# О ТЕРМОДИФФУЗИИ И КАЛИБРОВОЧНЫХ ПРЕОБРАЗОВАНИЯХ ДЛЯ ТЕРМОДИНАМИЧЕСКИХ ПОТОКОВ И СИЛ

Д.С. Голдобин

*Институт механики сплошных сред УрО РАН, Пермь, Российская Федерация*
*Пермский государственный национальный исследовательский университет, Пермь, Российская Федерация*

В работе рассматривается молекулярно-диффузионный транспорт в бесконечно разбавленных растворах при неизотермических условиях. Данное рассмотрение носит отчасти методический характер и мотивировано распространенностью искаженных интерпретаций уравнений термодинамического транспорта, записанных в терминах химического потенциала, при наличии градиента температуры. Уравнения транспорта содержат вклады, имеющие калибровочное происхождение, а именно, связанные с тем, что химический потенциал определен с точностью до слагаемого (AT+B) с произвольными константами A и B, где A — константа, с точностью до которой определена энтропия, а B — константа, с точностью до которой определена потенциальная энергия. Коэффициенты пропорциональности между термодинамическими силами и перекрестными термодинамическими потоками имеют вклады, связанные с необходимой инвариантностью по отношению к калибровочным преобразованиям — эти вклады не связаны с реальными физическими эффектами перекрестного транспорта. Представляемый анализ полагается на рассмотрение баланса энтропии и может подсказать многообещающие подходы для задачи об аналитическом вычислении константы термодиффузии из первых принципов. Кроме того, для разбавленных растворов обсуждается невозможность бародиффузии, понимаемой как диффузионный поток, создаваемый непосредственно градиентом давления. В литературе же «бародиффузией» часто называют дрейф под действием внешней потенциальной силы (например, в электростатическом или гравитационном поле), при котором в итоговых уравнениях сила, действующая на частицы, выражается через гидростатический градиент давления, который эта сила создает. Очевидно, интерпретация последнего как бародиффузии не вполне корректна и может спровоцировать ошибки при попытке учесть истинно бародиффузионные потоки.

*Ключевые слова:* разбавленные растворы, перекрестные термодинамические эффекты, термодиффузия

# ON THERMAL DIFFUSION AND GAUGE TRANSFORMATIONS FOR THERMODYNAMIC FLUXES AND FORCES

D.S. Goldobin

*Institute of Continuous Media Mechanics UB RAS, Perm, Russian Federation*
*Perm State University, Perm, Russian Federation*

We discuss the molecular diffusion transport in infinitely dilute liquid solutions under non-isothermal conditions. This discussion is motivated by an occurring misinterpretation of thermodynamic transport equations written in terms of chemical potential in the presence of temperature gradient. The transport equations contain the contributions owned by a gauge transformation related to the fact that chemical potential is determined up to the summand of form (AT+B) with arbitrary constants A and B, where constant A is owned by the entropy invariance with respect to shifts by a constant value and B is owned by the potential energy invariance with respect to shifts by a constant value. The coefficients of the cross-effect terms in thermodynamic fluxes are contributed by this gauge transformation and, generally, are not the actual cross-effect physical transport coefficients. Our treatment is based on consideration of the entropy balance and suggests a promising hint for attempts of evaluation of the thermal diffusion constant from the first principles. We also discuss the impossibility of the «barodiffusion» for dilute solutions, understood in a sense of diffusion flux driven by the pressure gradient itself. When one speaks of «barodiffusion» terms in literature, these terms typically represent the drift in external potential force field (e.g., electric or gravitational fields), where in the final equations the specific force on molecules is substituted with an expression with the hydrostatic pressure gradient this external force field produces. Obviously, the interpretation of the latter as barodiffusion is fragile and may hinder the accounting for the diffusion fluxes produced by the pressure gradient itself.

## 1. Введение

В раде задач механики многокомпонентных жидких смесей диффузия может как оказываться существенным осложняющим фактором, так и иметь решающее влияние на динамику системы [1–6]. Вместе с тем, в неизотермических системах молекулярная диффузия сопровождается перекрестным термодинамическим эффектом — термодиффузией [1], количественное влияние которой на динамику системы сопоставимо с влиянием молекулярной диффузии. В некоторых случаях термодиффузия может приводить к качественно новым явлениям в динамике сплошносредных систем [3–6].

В настоящей работе обсуждаются процессы молекулярно-диффузионного транспорта в бесконечно разбавленных жидких растворах при неизотермических условиях. Это обсуждение мотивировано распространенностью в литературе неточных интерпретаций уравнений термодинамического транспорта, записанных в терминах химического потенциала, в присутствии градиента температуры. Транспортные уравнения содержат вклады, имеющие калибровочное происхождение, а именно, связанные с тем, что химический потенциал $\mu$ определен с точностью до слагаемого $(AT+B)$ с произвольными константами $A$ и $B$. Здесь $A$ — постоянное слагаемое, с точностью до которого определена энтропия, $B$ — постоянное слагаемое, с точностью до которого определена потенциальная энергия. Сдвиг химического потенциа-ла на слагаемое ($AT + B$) ведет к изменению и формальной термодинамической силы, вызывающей молекулярно-диффузионный поток, $\nabla(-\mu/T)$. Поскольку термодинамические потоки при этом должны быть неизменными, очевидно, что изменение формального значения термодинамической силы должно компенсироваться. Эта компенсация происходит за счет изменения коэффициентов пропорциональности между термодинамическими силами и потоками для перекрестных эффектов. Коэффициенты перекрестных эффектов оказываются модифицированными калибровочным преобразованием, а не являются реальными физическими транспортными коэффициентами. Представляемый ниже анализ базируется на рассмотрении баланса энтропии (см., например, [1, 7–9]) и может подсказать перспективные подходы к решению задачи о вычислении константы термодиффузии из первых принципов.

Кроме того, будет разобран вопрос о невозможности эффекта бародиффузии для бесконечно разбавленных растворов, если этот эффект понимать естественным образом, как индуцирование диффузионного потока непосредственно градиентом давления. Как правило, когда в литературе (например, [1]) речь идет о бародиффузионных слагаемых, эти слагаемые описывают дрейф частиц в поле внешней потенциальной силы (электростатической или гравитационной), но в итоговых уравнениях сила, действующая на индивидуальные частицы, выражена через создаваемый ей макроскопический градиент гидростатического давления. Очевидно, такая интерпретация бародиффузии не вполне естественна и может привести к ошибкам при обращении к задаче о вычислении диффузионных потоках создаваемых именно непосредственно градиентом давления.

Положение дел в современной литературе (особенно в работах по аналитическому вычислению константы термодиффузии для жидких растворов) делает востребованным лаконичный, но самодостаточный разбор соотношения между калибровочными преобразованиями и перекрестными термодинамическими эффектами на прозрачном примере, где были бы очевидны физические интерпретации строго получаемых результатов.

## 2. Неизотермическая диффузия в поле внешней потенциальной силы

### 2.1. Общий термодинамический анализ

Выведем диффузионный поток растворенного вещества в неизотермическом растворителе в соответствии с базовым термодинамическом принципом Онзагера [7,8]. Рассмотрение ограничим случаем бесконечно разбавленных растворов. Вывод начнем с первого закона термодинамики для двухкомпонентных смесей:

$$dU = TdS - PdV + \sum_{j=1,2} \mu_j dN_j. \qquad (1)$$

Здесь использованы стандартные обозначения: $S$ — энтропия системы, $U$ — внутренняя энергия, $V$ — объем, $N_1$, $N_2$ и $\mu_1$, $\mu_2$ — количества частиц и химические потенциалы растворителя и растворяемого вещества, соответственно (далее индексы 1 и 2 всюду обозначают растворитель и растворяемое вещество, соответственно). Объемные концентрации частиц $n_j = N_j/V$. Рассматриваются разбавленные растворы: $n_2 \ll n_1$ (что типично, например, для растворов газов в жидкостях). В разбавленных растворах молекулы растворенного вещества практически не взаимодействуют друг с другом и, как следствие, можно пренебречь вариациями объема системы, связанными с перераспределением раствора. Математически это означает $v_1 N_1 + v_2 N_2 = V$, где $v_j$ — объем, приходящийся на одну частицу вещества $j$ в растворе. Полагаем жидкость несжимаемой: $V = \mathrm{const}$ и $v_{1,2} = \mathrm{const}$. Следовательно, для всех транспортных процессов будет справедливо условие $v_1 dN_1 + v_2 dN_2 = 0$, подразумевающее соотношение $dN_1 = -(v_2/v_1)dN_2$. Также будет полагаться, что для малых изменений плотности жидкости, связанных с тепловым расширением, это соотношение остается справедливо в ведущем порядке. Итогово, будет рассматриваться изменение $S$ как функции $U$, $n_1$ и $n_2$ для фиксированного объема $V$.

В терминах удельных величин $u = U/V$ и $s = S/V = s(u, n_1, n_2)$ для слабонеравновесной системы может быть записано

$$\frac{dS}{dt} = \frac{d}{dt}\int_V s\, dV = \int_V \frac{\partial s}{\partial t} dV = \int_V \left(\left(\frac{\partial s}{\partial u}\right)_{n_1,n_2} \frac{\partial u}{\partial t} + \left(\frac{\partial s}{\partial n_1}\right)_{u,n_2} \frac{\partial n_1}{\partial t} + \left(\frac{\partial s}{\partial n_2}\right)_{u,n_1} \frac{\partial n_2}{\partial t}\right) dV. \qquad (2)$$

Индексы при частных производных, в соответствии со стандартными правилами, обозначают переменные, остающиеся постоянными при дифференцировании. Из (1) можно получить

$$\left(\frac{\partial S}{\partial U}\right)_{V,N_1,N_2} = \frac{1}{T}, \quad \left(\frac{\partial S}{\partial N_1}\right)_{U,V,N_2} = -\frac{\mu_1}{T}, \quad \left(\frac{\partial S}{\partial N_2}\right)_{U,V,N_1} = -\frac{\mu_2}{T}.$$

Первая из приведенных производных идентична $(\partial s / \partial u)_{n_1,n_2} = 1/T$, две производные по $n_j$ могут быть скомбинированы следующим образом:

$$\left(\frac{\partial s}{\partial n_1}\right)_{u,n_2} dn_1|_V + \left(\frac{\partial s}{\partial n_2}\right)_{u,n_1} dn_2|_V = \frac{1}{V}\left(\left(\frac{\partial S}{\partial N_1}\right)_{U,V,N_2} dN_1|_V + \left(\frac{\partial S}{\partial N_2}\right)_{U,V,N_1} dN_2|_V\right) =$$

$$= \frac{1}{V}\left(\left(-\frac{\mu_1}{T}\right)dN_1|_V + \left(-\frac{\mu_2}{T}\right)dN_2|_V\right) = -\frac{1}{V}\left(\frac{\mu_2}{T} - \frac{v_2}{v_1}\frac{\mu_1}{T}\right)dN_2|_V = -\frac{\mu_2 - (v_2/v_1)\mu_1}{T}dn_2,$$

где использовано условие $dN_1 = (v_2/v_1)dN_2$. Отсюда,

$$\frac{dS}{dt} = \int_V \left(\frac{1}{T}\frac{\partial u}{\partial t} + \left(-\frac{\tilde{\mu}_2}{T}\right)\frac{\partial n_2}{\partial t}\right)dV, \tag{3}$$

где $\tilde{\mu}_2 = \mu_2 - (v_2/v_1)\mu_1$. Эффективный химический потенциал $\tilde{\mu}$ вводился также в работе [10] для термодинамического рассмотрения эффекта термодиффузии. Здесь и далее для различных термодинамических переменных $x_j$ используется обозначение $\tilde{x}_2 = x_2 - (v_2/v_1)x_1$. Частные производные по времени в уравнении (3) связаны с соответствующими потоками: $(\partial u / \partial t) = -\operatorname{div}\vec{J}_u$ и $(\partial n_2 / \partial t) = -\operatorname{div}\vec{J}_{n_2}$, где $\vec{J}_u$ и $\vec{J}_{n_2}$ — потоки величин $u$ и $n_2$, соответственно. Отсюда имеем

$$\frac{dS}{dt} = \int_V \left(-\frac{1}{T}\operatorname{div}\vec{J}_u + \frac{\tilde{\mu}_2}{T}\operatorname{div}\vec{J}_{n_2}\right)dV. \tag{4}$$

Для изолированной системы (системы, в которой потоки обращаются в ноль на границе) интегрирование по частям позволяет переписать последнее уравнение:

$$\frac{dS}{dt} = \int_V \left(\vec{J}_u \cdot \nabla \frac{1}{T} + \vec{J}_{n_2} \cdot \nabla\left(-\frac{\tilde{\mu}_2}{T}\right)\right)dV. \tag{5}$$

Для слабонеравновесной системы потоки линейны по градиентам:

$$\begin{pmatrix}\vec{J}_u \\ \vec{J}_{n_2}\end{pmatrix} = \mathbf{A} \cdot \begin{pmatrix}\nabla(1/T) \\ \nabla(-\tilde{\mu}_2/T)\end{pmatrix},$$

где $\mathbf{A}$ — матрица соответствующих транспортных коэффициентов. Положительность $dS/dt$ в уравнении (5) для любых состояний системы требует, чтобы матрица $\mathbf{A}$ имела только положительные собственные значения. В частности, отсутствие комплексных собственных значений требует, чтобы матрица $\mathbf{A}$ была симметрична:

$$\vec{J}_u = K\nabla\frac{1}{T} + \alpha n_2 \nabla\left(-\frac{\tilde{\mu}_2}{T}\right), \tag{6}$$

$$\vec{J}_{n_2} = \alpha n_2 \nabla\frac{1}{T} + \frac{Dn_2}{k_B}\nabla\left(-\frac{\tilde{\mu}_2}{T}\right), \tag{7}$$

где $(K/T^2)$ — коэффициент теплопроводности, $D$ — коэффициент молекулярной диффузии, $k_B$ — константа Больцмана, $\alpha$ — коэффициент, описывающий перекрестные потоки. Коэффициент $\alpha$ не следует считать тождественным константе термодиффузии [9]. Отметим, что выведенное уравнение (6) соответствует принципу Онзагера [7,8]. Теперь уравнение (5) принимает вид

$$\frac{dS}{dt} = \int_V \left(K\left(\nabla\frac{1}{T}\right)^2 - 2\alpha n_2 \nabla\frac{1}{T} \cdot \nabla\left(-\frac{\tilde{\mu}_2}{T}\right) + \frac{Dn_2}{k_B}\left(\nabla\frac{\tilde{\mu}_2}{T}\right)^2\right)dV. \tag{8}$$

Условие производства энтропии $dS/dt > 0$ для любого неравновесного состояния требует, чтобы квадратичная форма от градиентов $\nabla(1/T)$ и $\nabla(\tilde{\mu}_2/T)$ под интегралом в уравнении (8) была положительной. Это дает ограничение на значения коэффициентов: $KD > \alpha^2 k_B n_2$; последнее условие очевидно выполняется для разбавленных растворов, $n_2/n_1 \ll 1$, даже если $\alpha$ довольно велико по сравнению с коэффициентами теплопроводности и диффузии (например, $\alpha^2 > KDv_{\text{liq},1}/R$, где $v_{\text{liq},1}$ — объем 1 моля растворителя, $R$ — универсальная газовая постоянная).

### 2.2. Анализ системы с точки зрения микроскопической кинетики

Вышеприведенный вывод основан на общих принципах термодинамики. Для понимания смысла коэффициента $\alpha$ с точки зрения микроскопических процессов рассмотрим микроскопическое выражение для потока энергии $\vec{J}_u$, включающего кинетический перенос энергии молекулами и поток энергии за счет фононов $\vec{J}_{\text{Ph}}$,

$$\vec{J}_u = u_1 \vec{J}_{n_1} + u_2 \vec{J}_{n_2} + \vec{J}_{\text{Ph}}, \tag{9}$$

где $u_j$ — внутренняя энергия, приходящаяся на одну молекулу вещества $j$, $\vec{J}_{n_j}$ — поток соответствующих частиц. Поскольку $\vec{J}_{n_1} = -(v_2/v_1)\vec{J}_{n_2}$, то $u_1 \vec{J}_{n_1} + u_2 \vec{J}_{n_2} = \tilde{u}_2 \vec{J}_{n_2}$, где $\tilde{u}_2 = u_2 - (v_2/v_1)u_1$. Подстановка уравнения (7) в выражение для потока $\vec{J}_u$ дает

$$\vec{J}_u = \alpha n_2 \tilde{u}_2 \nabla \frac{1}{T} + \frac{D n_2}{k_B} \tilde{u}_2 \nabla\left(-\frac{\tilde{\mu}_2}{T}\right) + \vec{J}_{\text{Ph}}. \tag{10}$$

Фононный поток энергии $\vec{J}_{\text{Ph}}$ вызывается градиентом температуры и неоднородностью концентрации раствора, поскольку концентрация гостевых молекул влияет на локальный спектр фононов и их рассеивание:

$$\vec{J}_{\text{Ph}} = -\lambda_0 \nabla T - \chi T \nabla n_2 = -\lambda_0 \nabla T - \chi T n_2 \nabla \ln n_2. \tag{11}$$

Здесь $\lambda_0$ — теплопроводность жидкости без раствора (в самом деле, при $n_2 = 0$ полный поток энергии обращается в $-\lambda_0 \nabla T$), коэффициент $\chi$ может в общем случае зависеть от температуры, но не зависит от $n_2$ для разбавленных растворов. Для обращения к принципу Онзагера следует переписать поток энергии $J_u$ в терминах естественных термодинамических сил $\nabla(1/T)$ и $\nabla(-\tilde{\mu}_2/T)$ (см. уравнения (6) и (7)). Перепишем химический потенциал в следующем виде:

$$\mu_2 = k_B T \ln n_2 + \Delta\mu_2, \tag{12}$$

где $\Delta\mu_2 = \mu_2 - k_B T \ln n_2$ — сумма идеальной части химического потенциала, которая зависит от внутренних степеней свободы газовой молекулы, неидеальной части, описывающей взаимодействие с растворителем, и части, содержащей локальное давление. Важно, что для разбавленных растворов $\Delta\mu_2$ не зависит от $n_2$ (см., например, [10]). Для разбавленных растворов $|\nabla n_1|/n_1 \ll |\nabla n_2|/n_2$ и можно записать

$$\nabla \ln n_2 = \nabla \frac{\tilde{\mu}_2 - \Delta\tilde{\mu}_2}{k_B T} = -\frac{1}{k_B} \nabla\left(-\frac{\tilde{\mu}_2}{T}\right) - \frac{\partial}{\partial T}\left(\frac{\Delta\tilde{\mu}_2}{k_B T}\right) \nabla T. \tag{13}$$

Таким образом, поток энергии принимает вид

$$\vec{J}_u = \left(\lambda_0 T^2 - \chi T^3 n_2 \frac{\partial}{\partial T} \frac{\Delta\tilde{\mu}_2}{k_B T} + \alpha n_2 \tilde{u}_2\right) \nabla \frac{1}{T} + \frac{(D\tilde{u}_2 + \chi T) n_2}{k_B} \nabla\left(-\frac{\tilde{\mu}_2}{T}\right). \tag{14}$$

Сопоставляя последнее выражение для $\vec{J}_u$ с уравнением (6), можно видеть, что

$$K = \lambda_0 T^2 - \chi T^3 n_2 \frac{\partial}{\partial T} \frac{\Delta\tilde{\mu}_2}{k_B T} + \alpha n_2 \tilde{u}_2, \tag{15}$$

$$\alpha = \frac{D\tilde{u}_2 + \chi T}{k_B}. \tag{16}$$

Следовательно, поток раствора

$$\vec{J}_u = -Dn_2\left[-\left(\frac{\chi T}{k_B D} + \frac{\tilde{u}_2}{k_B}\right)\nabla\frac{1}{T} + \nabla\frac{\tilde{\mu}_2}{k_B T}\right] =$$

$$= -Dn_2\left[-\frac{\chi T}{k_B D}\nabla\frac{1}{T} + \nabla\frac{\tilde{\mu}_2 - \tilde{u}_2}{k_B T} + \nabla\frac{\tilde{u}_2}{k_B T}\right].$$

Используя теплоемкость частиц растворителя $c_{V,1}$ и гостевых частиц в растворе $c_{V,2}$, которые не зависят от $n_2$ для разбавленных растворов, можно записать $\tilde{u}_2 = \tilde{u}_{2,0} + \int^T \tilde{c}_{V,2}(T')dT' + \tilde{\phi}_2$, где $\tilde{u}_{2,0}$ не зависит от температуры и $\phi_j$ — потенциальная энергия молекулы вещества $j$ во внешнем поле (например, поле тяжести). Далее, для разбавленных растворов с пренебрежимым тепловым расширением $(\partial v_j/\partial T) \ll (v_j/T)$ — что, например, приблизительно верно для водных растворов инертных газов — может быть получено

$$\tilde{\mu}_2 = T\tilde{S}_{0,2} + T\int^T (\tilde{c}_{V,2}(T')/T')dT' + \tilde{\phi}_2,$$

где $\tilde{S}_{0,2}$ почти не зависит от температуры и давления. Окончательно, для разбавленных растворов, $|\nabla n_1|/n_1 \ll |\nabla n_2|/n_2$, уравнение (7) принимает вид

$$\vec{J}_{n_2} = -Dn_2\left[\frac{\nabla n_2}{n_2} + \frac{\chi}{k_B TD}\nabla T + \frac{\nabla \tilde{\phi}_2}{k_B T}\right]. \tag{17}$$

Для случая гравитационного поля $\tilde{m}_2 \vec{g} = -\nabla\tilde{\phi}_2$, и последнее уравнение переписывается в виде

$$\vec{J}_{n_2} = -Dn_2\left[\frac{\nabla n_2}{n_2} + \alpha_T \frac{\nabla T}{T} - \frac{\tilde{M}\vec{g}}{RT}\right]. \tag{18}$$

Здесь $\tilde{M} = M_2 - (v_2/v_1)M_1$, $M_j$ — молярная масса, и

$$\alpha_T = \frac{\chi}{k_B D} \tag{19}$$

— константа термодиффузии.

В теоретической работе [10] представлен альтернативный подход к вычислению константы термодиффузии. В частности, вычислен изотопический вклад в коэффициент термодиффузии

$$\alpha_{T,\text{isot}} = \frac{3}{4}\ln\frac{M_2 v_1}{M_1 v_2}, \tag{20}$$

который оказывается независящим от температуры. Выражение (20) дает оценку, заниженную приблизительно в 3 раза по сравнению с экспериментальными данными [11].

### 3. Заключение

Заслуживает особого акцента то, что для неизотермических систем в термодинамических потоках (6) и (7) слагаемое $\nabla(-\tilde{\mu}_2/T)$ описывает не чисто фиковскую диффузию, как и слагаемое с $\alpha$ не является чисто термодинамическим перекрестным эффектом. Эти слагаемые взаимно переплетены, и слагаемое с $\alpha$ может присутствовать даже в отсутствии перекрестного термодинамического эффекта, будучи полностью связанным с калибровочным преобразованием, обусловленным инвариантностью химического потенциала по отношению к добавлению ($AT + B$) с произвольными константами $A$ и $B$.

Имея соотношения между $\alpha_T$ и $\chi$ (уравнение (19)), можно рассмотреть нестандартный путь подхода к теоретическому вычислению константы термодиффузии на основе рассмотрения не потока молекул, а по-тока тепловой энергии $J_\mathrm{Ph}$ через объём с поддерживаемым однородным полем температуры и градиентом концентрации частиц.

На начальном этапе представленного вывода были исключены вклады в производство энтропии за счёт изменения объёма, $(\partial S/\partial V)_{U,N_1,N_2}$, которые внесли бы в систему термодинамические потоки, с соответствующей термодинамической силой $\nabla(P/T)$. Эта сила связана со сжимаемостью и была исключена по причине того, что изменение объёма бесконечно разбавленных растворов за счёт перераспределения частиц раствора исчезающе мало. Перекрёстный эффект, связанный с $\nabla(P/T)$, является бародиффузией, которая, следовательно, исчезает для разряжённых растворов, рассматриваемых в настоящей работе. Однако, нередко в литературе (например, [1]) слагаемое, описывающее дрейф в поле внешней потенциальной силы, т.е. последнее слагаемое в уравнении (17), переписывается в терминах гидростатического градиента давления, создаваемого внешней силой. Например, в [1] массовая сила (ускорение свободного падения) $\vec{g}$ заменяется на $\rho^{-1}\nabla P$. После такой замены все слагаемые, содержащие градиент давления, относятся к «бародиффузии». Однако, имеет смысл быть осторожными с такой трактовкой бародиффузии и различать потоки, создаваемые внешними силами, и потоки, вызываемые термодинамической силой $\nabla(P/T)$, полагая только последние относящимися к эффекту бародиффузии. Интерпретация первых как бародиффузии не вполне «физична» и может приводить к ошибкам при обращении к задаче о вычислении диффузионных потоков, создаваемых собственно градиентом давления.

Теоретическое вычисление константы термодиффузии из первых принципов востребовано в связи с отсутствием экспериментальных данных относительно величины константы для слабо растворимых газов, таких как метан, углекислый газ и т.п., и важностью термодиффузии водных растворов этих газов для процессов формирования природных депозитов гидрата метана и захоронения промышленных выбросов углекислого газа в морских донных отложениях [7–9]. В целом, перекрёстные термодинамические эффекты являются важными осложняющими факторами в сплошносредных системах, динамика которых существенным образом зависит от процессов молекулярной диффузии (например, [10]).

Для дополнительного чтения по смежным вопросам заинтересованным читателям может быть рекомендована монография [9].

**Сведения об авторах**

**a) Голдобин Денис Сергеевич, к.ф.-м.н.**
**b) Goldobin Denis Sergeevich**
**c) Федеральное государственное бюджетное учреждение науки Институт механики сплошных сред Уральского отделения Российской академии наук**
**d) Старший научный сотрудник, руководитель Группы динамики геологических систем**
**e) Denis.Goldobin@gmail.com**